\documentclass[journal]{IEEEtran}

\usepackage{amsmath,amsfonts,bm}
\usepackage{array}
\usepackage{textcomp}
\usepackage{stfloats}
\usepackage{url}
\usepackage{verbatim}
\usepackage{graphicx}
\usepackage{cite}
\usepackage{amssymb}
\usepackage{CJK}
\usepackage[ruled,linesnumbered, noend]{algorithm2e}
\usepackage{booktabs}
\usepackage{diagbox}

\ifCLASSOPTIONcompsoc
\usepackage[caption=false, font=footnotesize, labelfont=sf, textfont=sf,subrefformat=parens]{subfig}
\else
\usepackage[caption=false, font=footnotesize,,subrefformat=parens]{subfig}
\usepackage{makecell}
\usepackage{multirow}
\usepackage{xcolor}
\usepackage{tabularx}
\usepackage{bbding}
\usepackage{threeparttable}
\usepackage{color}

\begin{document}

\title{Generative AI for Energy Harvesting Internet of Things Network: Fundamental, Applications, and Opportunities}

\author{
        Wenwen~Xie,
        Geng~Sun,    
        Jiahui~Li,
        Jiacheng~Wang,
        Hongyang~Du,
        Dusit~Niyato,~\IEEEmembership{Fellow,~IEEE}, \\
        Octavia A. Dobre,~\IEEEmembership{Fellow,~IEEE}
        \IEEEcompsocitemizethanks
        {
        \IEEEcompsocthanksitem W.~Xie and J.~Li are with the College of Computer Science and Technology, Jilin University, Changchun 130012, China~(e-mail: xieww22@mails.jlu.edu.cn, lijiahui0803@foxmail.com).
         \IEEEcompsocthanksitem G.~Sun is with the College of Computer Science and Technology, Jilin University, Changchun 130012, China, and also with the College of Computing and Data Science, Nanyang Technological University, Singapore 639798 (e-mail: sungeng@jlu.edu.cn).
        \IEEEcompsocthanksitem J.~Wang, H.~Du and D.~Niyato are with the College of Computing and Data Science, Nanyang Technological University, Singapore 639798 (e-mail: jiacheng.wang@ntu.edu.sg, hongyang001@e.ntu.edu.sg, dniyato@ntu.edu.sg).
        \IEEEcompsocthanksitem Octavia A. Dobre is with the Faculty of Engineering and Applied Science, Memorial University, Canada (e-mail: odobre@mun.ca).
        }
 }

\maketitle

\begin{abstract}
\par Internet of Things (IoT) devices are typically powered by small-sized batteries with limited energy storage capacity, requiring regular replacement or recharging. To reduce costs and maintain connectivity in IoT networks, energy harvesting technologies are regarded as a promising solution. Notably, due to its robust analytical and generative capabilities, generative artificial intelligence (GenAI) has demonstrated significant potential in optimizing energy harvesting networks. Therefore, we discuss key applications of GenAI in improving energy harvesting wireless networks for IoT in this article. Specifically, we first review the key technologies of GenAI and the architecture of energy harvesting wireless networks. Then, we show how GenAI can address different problems to improve the performance of the energy harvesting wireless networks. Subsequently, we present a case study of unmanned aerial vehicle (UAV)-enabled data collection and energy transfer. The case study shows distinctively the necessity of energy harvesting technology and verify the effectiveness of GenAI-based methods. Finally, we discuss some important open directions.

\end{abstract}


%
\section{Introduction}

\par With the rapid development of the Internet of Things (IoT), it has spawned a number of domains such as smart transportation, healthcare and environmental monitoring~\cite{Khanna2020}. In these domains, a large number of smart devices or sensors are typically deployed to collect surrounding environment information for further analysis and decision-making, which is crucial for the intelligent control. However, these IoT devices are typically energy-constrained in practice, which greatly limits their availability. It is worth noting that traditional methods of manually replacing batteries and wired charging increase the maintenance effort and operating costs. Therefore, it is especially important to find a new solution to provide IoT devices with continuous energy supply to ensure the reliable network operation.

\par Energy harvesting technology that collect energy from the environment to power or store for future use can provide a promising solution for IoT networks~\cite{Lu2015}. One feasible method is to harvest energy from renewable natural sources such as solar, wind, and vibrational energy. The benefit of this method is energy friendliness, which reduces the burden on the environment and promotes the development of green technology and sustainability. However, the supply of renewable natural energy can be affected by weather conditions and time, which may lead to instability of the network. Unlike renewable natural energy sources, radio frequency (RF) energy is more stable, making RF energy transmission and collection technology an alternative method for powering the next generation of wireless networks.

\par RF energy can be broadly classified into two types. One is ambient RF, which refers to the RF signals that are not intended for RF energy transfer but for other purposes such as data transmission. The other one involves dedicated RF sources that can be used to deliver energy to network nodes where a reliable energy supply is necessary. Because the dedicated RF source is fully controllable, it is suited for for supporting network applications with quality of service (QoS) requirements. Moreover, the dedicated RF sources can include mobile units like unmanned aerial vehicles (UAVs) that are capable of periodically moving and delivering RF energy to network nodes, thereby improving the quality of energy transfer service.~\cite{Liu2022}.


\par Note that optimizing energy harvesting wireless networks from the perspectives of resource allocation and network deployment is crucial for enhancing overall network performance. However, conventional optimization strategies, such as convex optimization, have notable limitations. Firstly, traditional optimization lacks adaptability, which makes it struggle to effectively address dynamic optimization in energy harvesting wireless networks. Secondly, due to the inherent uncertainties and unpredictability in energy harvesting wireless networks, handling situations where network states or parameters are unknown poses a significant challenge for traditional optimization. Therefore, identifying a more suitable technique for energy harvesting wireless networks is of paramount importance.

\par With the recent popularity of ChatGPT and Sora, generative artificial intelligence (GenAI) technology has garnered widespread attention, particularly in the realm of multimedia content generation. It is worth noting that the capabilities of GenAI can be attributed to its following characteristics:

\begin{itemize}
    \item Data Augmentation: GenAI can generate highly similar and authentic data by learning training samples, which helps to alleviate data scarcity. For example, GenAI can produce more dataset to train UAV trajectory optimization.
    \item Latent Representation Learning: GenAI can learn the distribution of training data in order to mine the latent features and patterns of the data, which is conducive to dimensionality reduction of high-dimensional data. For instance, GenAI can extract  most important channel features of UAV network environments.
    \item Knowledge Transfer: GenAI can transfer the learned knowledge among domains by unifying the distribution of newly generated data to the given data. For example, GenAI can apply the energy harvesting knowledge learned in urban environments to other environments.
\end{itemize}

\par Thanks to these characteristics, GenAI is highly suitable for energy harvesting wireless networks. For instance, GenAI can predict the effects of different antenna configurations on energy harvesting and network coverage, thereby identifying the optimal antenna design solutions~\cite{nando2024enhancing}. Moreover, the robust analytical capability of GenAI can be employed to estimate channel conditions, which is crucial for energy harvesting wireless networks~\cite{zargari2024deep}. Although integrating GenAI into energy harvesting wireless networks offers significant advantages, there exist several issues that require further investigation. 

\par Motivated by this, we provide a comprehensive tutorial to present important applications of GenAI on energy harvesting wireless networks. The contributions of our work can be summarized as follows:
\begin{itemize}
    \item We first present fundamentals of common GenAI models. Subsequently, we discuss energy harvesting technologies based on renewable natural source and RF energy source with their issues that can be addressed by GenAI technologies. Furthermore, we demonstrate the basic architecture of an energy harvesting wireless network. 
    \item We explore how GenAI can address an optimization problems in energy harvesting wireless networks from multiple perspectives, \textit{i.e.}, channel estimation, relay topology design, antenna design, secure cognitive radio communication, and renewable energy generation forecasting.
    \item We introduce a case study on UAV energy harvesting and information transfer. Experimental results demonstrate the effectiveness of GenAI in energy harvesting wireless networks. 
\end{itemize}

\section{Overview of Energy Harvesting}
\par Energy harvesting wireless networks represent a significant advancement for wireless communication, where devices can self-sustain by harvesting energy from their environment.

\subsection{Renewable Natural Energy Harvesting}
\par Renewable natural energy refers to energy sources that can be naturally replenished and sustainably utilized~\cite{ahmadi2019renewable, Sah2020}. Currently, renewable natural energy is a significant energy source for rechargeable wireless devices, and common types of renewable energy are as follows:

\subsubsection{Solar Energy} Solar energy harvesting typically involves converting sunlight into electricity using photovoltaic cells, which is highly effective in sunny outdoor environments. Consequently, solar energy harvesting is widely applied in remote sensing, environmental monitoring, and agricultural IoT systems.

\subsubsection{Wind Energy} Wind energy harvesting typically employs wind turbines to convert the kinetic energy of wind into electricity, which is particularly suitable for areas with abundant wind energy resources. Moreover, wind power generation devices can be installed near wireless network base stations or nodes to provide them with sustained power supply.

\subsubsection{Other Energy} Other forms of renewable energy also hold potential in energy harvesting for wireless networks. For example, thermoelectric generators convert temperature differences into electricity, while piezoelectric materials generate electricity through mechanical vibrations. These energy sources are commonly found in industrial environments and can provide power support for industrial wireless sensors widely.

\par The most notable advantage of renewable natural energy is its environmental friendliness. However, renewable natural energy still has limitations, such as geographic constraints or instability due to environmental conditions. The emergence of RF energy harvesting offers a new approach to alleviate such limitations, especially dedicated RF sources.

\subsection{RF Energy Harvesting}
\par RF energy harvesting allows wireless devices to harvest energy from RF signals for their information processing and transmission. This technique becomes a promising solution for energy-constrained wireless networks. Note that the RF sources can be classified into two types, \textit{i.e.}, dedicated RF sources and ambient RF sources.
\begin{itemize}
    \item Dedicated RF Sources: Dedicated RF sources are specifically designed and implemented to provide RF energy for energy harvesting purposes. These sources are often tailored to deliver consistent and predictable energy output at certain frequencies and power levels to ensure efficient energy transfer to the network devices.
    \item Ambient RF Sources: Ambient RF sources refer to the naturally occurring or existing RF signals in the environment, which are not specifically designed for energy harvesting. These sources include RF emissions from various communication and broadcasting systems, such as TV, radio towers, and WiFi routers. It is worth noting that ambient RF sources can be also regarded as renewable energy sources.
\end{itemize}

\par Both dedicated RF sources and ambient RF sources play an important role in RF energy harvesting wireless networks. Specifically, dedicated RF sources provide efficient and controllable energy transfer, and they are suitable for the applications that require a stable energy supply. Moreover, ambient RF sources utilize existing RF signals to provide a flexible energy harvesting method for low-power devices, and they are suitable for a wide range of urban and portable device applications.

\begin{figure}
    \centering
    \includegraphics[width=\linewidth]{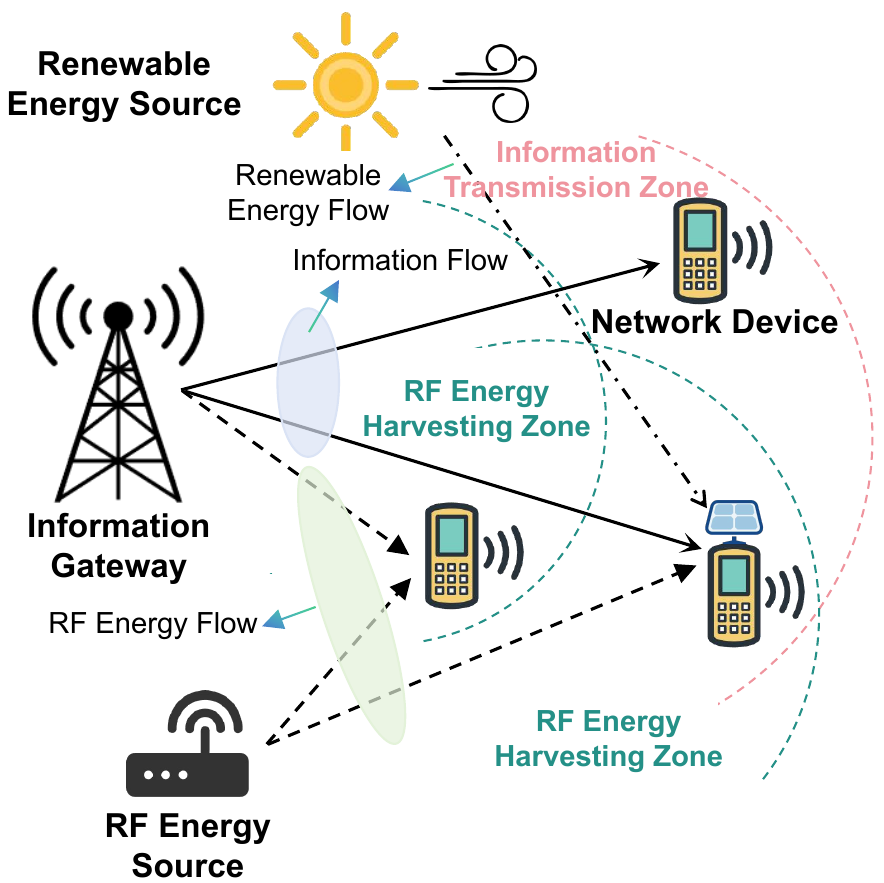}
    \caption{Energy Harvesting Architecture. The information gateway is responsible for data transmission with the network devices. The energy transmitter is responsible for delivering energy to the network devices, where the energy sources include renewable natural energy and RF energy.}
    \label{fig:energy harvesting architecture}
\end{figure}

\subsection{Energy Harvesting Architecture}
\par As shown in Fig.~\ref{fig:energy harvesting architecture}, the general 
infrastructure-based architecture of energy harvesting wireless network consists of the following three components:
\begin{itemize}
    \item Information Gateway: Information gateways, such as base stations and relay devices, can serve as communication hubs in energy harvesting wireless networks. Moreover, they are responsible for managing and coordinating the communication processes across an entire network. Information gateways typically have continuous and fixed power supplies and can provide power support to wireless devices.
    \item Energy Source: The energy source acts as the energy provider in the energy harvesting wireless network, supplying the necessary wireless power to network nodes. Note that, in some special cases, the information gateway and the RF energy source can coexist on the same device.
    \item Network Nodes/Devices: Network nodes refer to user devices within the network, responsible for executing specific application tasks. They can obtain sustainable power supplies from the energy transmitter to extend their operational time, and they can also communicate with the information gateway.
\end{itemize}
\par Note that infrastructure-less energy harvesting network has a similar architecture to the infrastructure-based one shown in Fig.~\ref{fig:energy harvesting architecture}, except that the network nodes communicate directly with each other.

\section{GenAI-enabled Energy Harvesting Network}
\label{sec:GenAI-enabled Energy Harvesting Network}
\par In this section, we first present several common GenAI models. Subsequently, the reasons for using GenAI to optimize energy harvesting are elaborated. Moreover, we introduce how GenAI can address an optimization problems in energy harvesting wireless networks.

\subsection{Overview of GenAI Models}


\par The emergence of GenAI has led AI technology to a whole new stage and brought new perspectives to a number of areas, including media content generation, system design, and wireless network optimization. In the following, several common GenAI models are described in detail.

\subsubsection{Generative Adversarial Network (GAN)}
\par GANs consist of two neural networks: a generator and a discriminator. The generator creates artificial data from random noise, while the discriminator attempts to distinguish between real and artificial data. Through adversarial training, the generator network generates high-quality data. GANs excel at generating highly realistic data but suffer from training instability and mode collapse that the generator keeps producing limited similar data patterns. GANs are primarily used in image generation, data augmentation, and network optimization.

\subsubsection{Diffusion Model}
\par Diffusion models gradually add noise to data through a forward diffusion process until it becomes pure noise, and then denoise it to recover the data through a reverse diffusion process. Note that the reverse process can use neural networks to denoise step-by-step. Diffusion models can produce high-quality data, making them suitable for image and speech generation. Moreover, diffusion models can be utilized to improve the policy network of deep reinforcement learning (DRL), which has been applied to several domains and has shown superior performance.

\subsubsection{Variational Autoencoder (VAE)}
\par VAEs consist of an encoder and a decoder. The encoder maps input data to a probabilistic distribution of latent variables, while the decoder reconstructs the data from these variables. VAEs are trained by minimizing reconstruction error and the Kullback Leibler (KL) divergence between the latent distribution and a standard normal distribution. They excel at creating continuous latent spaces and explicitly modeling generation probabilities, making them suitable for image generation and anomaly detection.

\subsubsection{Transformer}
\par The Transformer model consists of multiple stacked encoder and decoder layers, each containing self-attention mechanisms and feed-forward neural networks. The encoder transforms input sequences into context-aware representations, while the decoder generates the next sequence element based on the output and the previous output of the decoder. Transformers handle long-range dependencies well, offer high training efficiency, and have a flexible structure, making them widely used in natural language processing tasks such as machine translation and text generation.

\subsection{Reasons for GenAI on Energy Harvesting Optimization}
\par Most current work on energy harvesting wireless network focuses on how to design energy harvesting devices or systems from hardware perspectives~\cite{Guo2018,8322138,6206429}. It is equally important to improve the performance of the entire energy harvesting wireless network by optimizing the allocation of certain key resources. In this case, traditional methods such as evolutionary algorithm and convex optimization have the following limitations in energy harvesting wireless networks:
\begin{itemize}
    \item Poor Real-time Performance: In dynamic environments, traditional optimization methods (\textit{e.g.}, evolutionary algorithm) are difficult to quickly respond and adapt to environmental changes, thus affecting real-time performance optimization.
    \item Strong Model Dependency: Convex optimization requires an accurate mathematical model, and the uncertainty and dynamic changes in the actual network environment make it difficult to describe the model accurately.
\end{itemize}
\par Note that GenAI has powerful analyzing ability, generating ability, and dynamic environment adaptability, which provides new approaches to solve the limitations above. Therefore, we demonstrate how to improve the performance of an energy harvesting wireless network from different perspectives by using GenAI techniques.

\subsection{Roles of GenAI in Energy Harvesting IoT Networks}

\subsubsection{Channel Estimation}

\par The recent proliferation of connected devices has led to a substantial surge in spectrum utilization and energy consumption. As a solution, ambient backscatter communication (AmBC) has garnered considerable attention. Note that an AmBC node needs to harvest energy to supply its backscatter operations, \textit{i.e.}, circuit energy consumption. In this case, channel estimation is crucial in AmBC, since it can help to optimize energy utilization. Specifically, by accurately estimating the channel state, key resources such as frequency allocation can be dynamically adjusted to extend the working life of wireless devices. However, traditional channel estimation methods, including the techniques such as pilot-based estimators, blind estimators, and semi-blind estimators, typically face several issues such as requiring extensive prior knowledge, being limited by scenario changes, and exhibiting sensitivity to model matching. Therefore, advanced machine learning algorithms for channel estimation are of great interest.


\par In this context, a novel channel estimation algorithm based on conditional GAN (CGAN) for AmBC was proposed~\cite{zargari2024deep}. The CGAN estimator can estimate the wireless channel status and quality by accurately capturing complex features and utilizing stronger reflection paths. This capability allows the CGAN estimator to adapt well to the changes in the relative strength of reflected signals, resulting in superior channel estimation performance. Specifically, the proposed channel estimation method based on CGAN applies specific channel state information (CSI) as a condition to GAN, and the generator network generates synthetic channel data that closely resembles real-world conditions. This synthetic data in turn facilitates accurate estimation of AmBC channel coefficients.

\par The performance of the proposed CGAN estimator is evaluated by comparing it with the well-known methods such as least squares (LS), minimum mean-squared error (MMSE), blind, deep residual learning denoiser (CRLD), and convolutional neural network (CNN). The results show that CGAN channel estimator excels in normalized MSE (NMSE), even under challenging low signal-to-noise ratio (SNR) conditions. Moreover, at an SNR of 5 dB, the CGAN channel estimator consistently surpasses existing CE solutions, achieving an 82\% improvement over CRLD and MMSE estimators.

\subsubsection{Relay Topology Design}
\par Relay-based structures are particularly suitable for energy harvesting wireless networks. First, cooperative relaying techniques enhance network efficiency and reliability by using relay nodes to overcome fading and attenuation. Second, energy-rich nodes can transmit data on behalf of energy-constrained nodes to maximize overall network efficiency. Therefore, through proper deployment and energy management of relay nodes, the network can address energy harvesting uncertainties, extend node lifetimes, and improve data transmission reliability.

\par The authors in~\cite{Chung2023} considered an uplink transmission scenario in an relay-based IoT sensor network, where each wireless node self-powers by harvesting energy from power beacons (PBs) and stores remaining energy in a limited-capacity rechargeable battery. The authors aimed to optimize the topology matrix among nodes to maximize the transmission bits of all nodes. Given the lack of datasets and the nonlinear nature of the optimization problem, the authors proposed an unsupervised relay topology algorithm based on a VAE. Specifically, a random latent vector is input into the VAE, and subsequently the analytical and generative capabilities of the VAE decoder are utilized to generate the topology matrix. To evaluate quality of the topology matrix generated by the VAE, the authors introduced a packet-tracking evaluation method inspired by ray-tracing techniques. This method can accurately evaluate the output of the VAE, providing appropriate guidance for VAE training. 

\par The VAE-based topology approach was compared with traditional schemes such as the direct connect scheme, minimum spanning tree (MST) scheme, and greedy scheme, as well as the optimal brute-force solution. The experiment results indicate that the MST scheme exhibits insufficient adaptability to various IoT network configurations, and the brute-force search algorithm has excessively long computation times. In contrast, the VAE-based topology approach demonstrates better performance than all other considered approaches across all node counts and network configurations, being the closest to the optimal solution.


\subsubsection{Antenna Design}
\par Antennas are the core components of energy harvesting devices, which means that the antenna design is critical in energy harvesting wireless networks. By adjusting the antenna design in terms of frequency selectivity, directivity, and structure, we can optimize signal reception and minimize energy loss and interference. Consequently, an ideal and well-considered antenna design can maximize the ability of devices to harvest RF energy from the environment. 

\par The authors in~\cite{nando2024enhancing} investigated to enhance the RF energy harvesting and power transfer with GAN-optimized antenna design. Inspired by the quasi-Yagi antenna (QYA) model, the authors aimed to develop a cube-shaped three-dimensional (3D) antenna optimized for 915 MHz by analyzing various 2D QYA configurations. First, to enhance robustness of the analysis, the authors employed GAN to expand the dataset. Specifically, the generator network generates synthetic QYA data, while the discriminator network engages in adversarial training to optimize the performance of the generator network. Moreover, GAN was also used to optimize the design of the 3D QYA. A well-trained GAN can effectively capture the complex relationships between QYA parameters and QYA structure, generating target designs based on specified criteria. To strengthen this process, the authors integrated the GAN with the CST Studio Suite for virtual performance testing of GAN-generated designs to enable timely adjustments of the GAN model, thereby enhancing energy capture from ambient and dedicated RF sources. 

\par The experimental results show that the minor deviation between the GAN and CST Studio Suite results is within an acceptable range, underscoring the reliable performance of the antenna. Moreover, compared to other 2D QYA designs, the GAN-designed 3D antenna exhibits superior energy harvesting capability.

\subsubsection{Secure Cognitive Radio Communication}
\par Cognitive radio, through dynamic spectrum management and spectrum sharing, effectively alleviates spectrum scarcity issues, ensuring efficient network operation. Given that cognitive radio devices frequently engage in spectrum sensing and dynamic adjustments, traditional battery power typically fails to meet their long-term operational needs. Energy harvesting can provide continuous power support for these devices. Moreover, the benefit of energy harvesting can be further enhanced by optimizing the allocation of resources in the energy harvesting network.

\par In this case, the authors in~\cite{Lin2024} developed the GAN-based DRL algorithm to enhance the physical security in energy harvesting cognitive radio system. In the considered system, the communication security of the secondary user (SU) can be tampered by eavesdroppers, prompting the deployment of a cooperative jammer to secure wireless transmissions. Notably, both the SU and the jammer can harvest energy from the received RF signals emitted by the primary transmitter (PT), with batteries capable of storing energy for extended periods. The authors employed deep Q-network (DQN) algorithm to maximize the average minimum secrecy rate and minimize the average maximum secrecy outage probability (SOP). To better estimate the state-action values, GAN was introduced to learn the optimal state-action value distribution for DRL. Specifically, the generator network $G$ is responsible for creating the estimated state-action value distribution, while the target generator network $G^{'}$ produces the target state-action value distribution. The role of the discriminator network $D$ is to differentiate between the values produced by $G$ and those from $G^{'}$.

\par The experimental results indicate that, compared to the case with one eavesdropper, the secrecy rate achieved by DQN decreases by approximately 78\% and the SOP increases by 67\% when there are ten eavesdroppers. This suggests that the DQN approach is highly sensitive to the number of eavesdroppers. In contrast, the GAN-powered DQN reduces the secrecy rate by 20\% and improves the SOP by 67\%. Moreover, the maximum average SOP achieved with GAN-powered DQN closely matches the theoretical value of the closed expression, which validates the effectiveness of resisting eavesdropping.

\subsubsection{Renewable Energy Generation Forecasting}
\par Since the renewable energy is an intermittent generation, it is subject to external factors such as weather variations and changes in user behavior, which often bring new challenges to the energy harvesting solutions. In this context, accurate forecasting of power generation in advance is essential to improve the operational efficiency of the energy harvesting systems, which can be used in wireless networks.

\par A VAE-based method for solar power generation forecasting was proposed~\cite{kaur2023vae}. VAE can effectively reduce the dimensionality of parameters while retaining important data features, which is highly beneficial for enhancing the computational efficiency and generalization capability of the prediction model. The proposed VAE-based process involves mapping high-dimensional data to a low-dimensional latent space and then reconstructing the data from the latent space. Therefore, by optimizing the reconstruction error and the probability distribution of latent variables, VAE can learn a low-dimensional representation of the data. Building on the VAE, a Bayesian bidirectional long short-term memory (BiLSTM) network was incorporated for time series prediction. BiLSTM can capture temporal dependencies and bidirectional information in the data, further improving prediction accuracy. 

\par Compared with benchmark methods, the VAE-based approach shows significant improvement in parameter reduction. Specifically, in six months of solar power generation data, the number of weight parameters was reduced from 764224 to 2022, achieving a quantized parameter reduction of 97.35\% and a computation time improvement of 37.93\%.

\subsection{Lesson Learned}
\par From the applications above, we can find that GenAI improves the performance of the energy harvesting IoT networks from the following perspectives.

\begin{itemize}
    \item Data Augmentation: Leveraging the generative capabilities of GenAI, it can be used to produce data that closely resembles the training samples, thereby addressing the issue of dataset scarcity~\cite{nando2024enhancing}.
    \item Solution Exploration: GenAI can leverage its powerful learning and reasoning capabilities to find solutions for optimization problems~\cite{zargari2024deep, Chung2023}. Additionally, GenAI can be employed to enhance other algorithms, such as DRL, to improve their performance~\cite{Lin2024}.
    \item Dimensionality Reduction: GenAI can extract the latent representation variables of data to perform dimensionality reduction while retaining essential information~\cite{kaur2023vae}.
\end{itemize}

\begin{figure*}
    \centering
    \includegraphics[width=\linewidth]{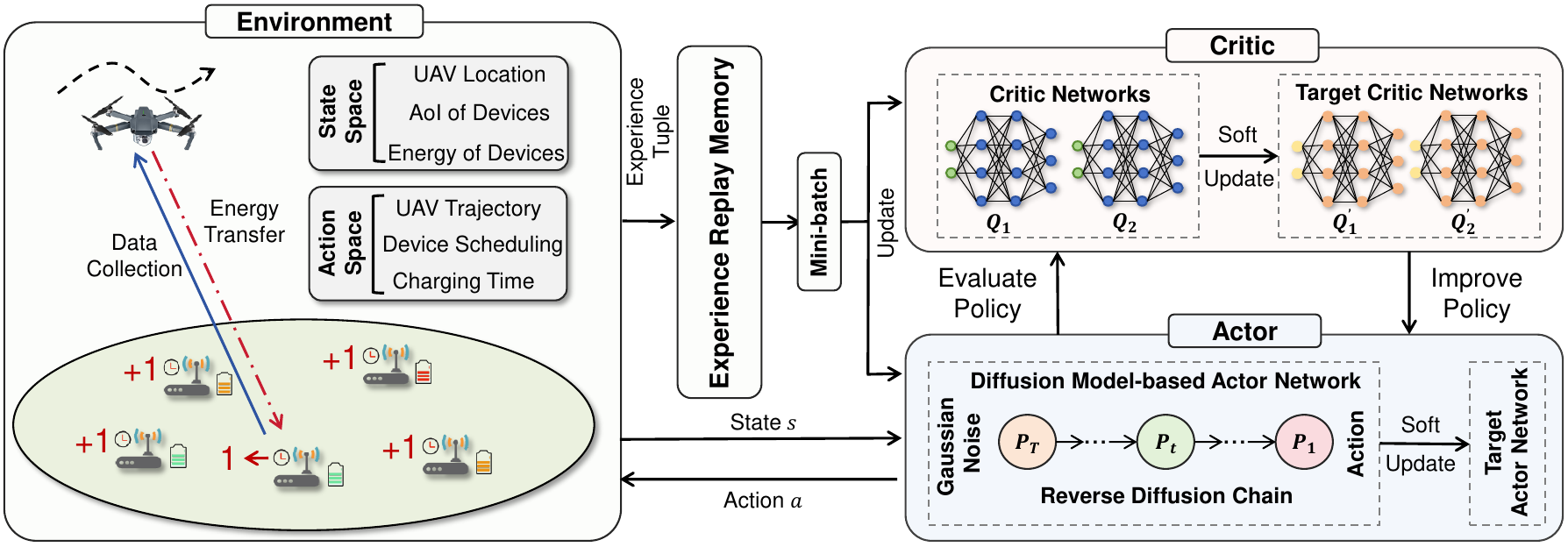}
    \caption{The framework of case study. The UAV is dispatched to charge a set of IoT devices and collect data from them to minimize AoI. Diffusion model is used to improve the actor network of TD3 algorithm to generate high quality decisions.}
    \label{fig:case study}
\end{figure*}

\section{Case Study: GenAI-enabled Energy Harvesting Wireless Network}

\par In this section, we present a case study to evaluate the performance of GenAI on energy harvesting wireless network. 

\subsection{Scenario Description}
\subsubsection{Motivation}
\par To improve online data analysis and decision-making in IoT networks, it is essential to effectively gather the latest information produced by IoT devices. To precisely measure information freshness, the concept of Age of Information (AoI) has been introduced~\cite{Hu2021}. AoI is defined as the duration from when the most recent update packet is created at the source until it is received at the destination. Note that enhancing the power supply for energy-constrained IoT devices is essential for improving AoI performance in IoT networks. However, traditional methods of energy supply, such as battery replacements or wired connections, are impractical and labor-intensive. Note that using UAVs for energy transmission has emerged as a promising solution to overcome the energy limitations of IoT devices. UAVs offer flexibility and mobility, allowing them to reach remote areas that traditional power sources might not be able to access. Therefore, UAVs can provide stable energy transfer in an on-demand basis, reducing downtime for IoT devices and ensuring the freshness of data collection and continuity of communication.

\subsubsection{Problem Formulation}

\par As shown in Fig.~\ref{fig:case study}, multiple IoT devices equipped with energy buffers are randomly distributed within the target area to sense surrounding environment. A UAV is moving within the target area, responsible for charging the IoT devices to ensure their activity and receiving data from the IoT devices. Note that only one IoT device can upload data during each time slot. We aim to minimize the AoI by optimizing the trajectory of UAV, the scheduling of IoT devices, and the charging time.

\subsection{Proposed Design}

\par Due to the mobility of the UAV, our considered scenario involves highly dynamic nature, which makes the traditional optimization methods difficult for handling such problems. Notably, DRL enables agents to adapt their behavior quickly through trial-and-error learning, thus effectively adapting to the changing environment of UAV-assisted IoT systems. Among the many DRL algorithms, the twin delayed deep deterministic policy gradient (TD3) algorithm stands out for its excellent stability~\cite{Fujimoto2018}.

\par To further enhance the performance of the TD3 algorithm, we propose a diffusion model-based TD3 (DTD3) algorithm, where the inverse process of the diffusion model is used to improve the actor network of TD3. Specifically, instead of the traditional Multi-Layer Perception (MLP), the actor network can be regarded as a denoiser, which starts from Gaussian noise and progressively recovers the optimal decision solution according to the environment conditions.

\subsection{Simulation Settings}

\begin{itemize}
    \item Environment and Tools: The server used for the simulation is equipped with an Intel i9 13900K CPU and an NVIDIA RTX 4090 GPU, and the RAM is a 32 GB DDR4. Moreover, the simulation code is implemented by Pytorch.
    \item Simulation Parameters: The critic network and the denoising network of the diffusion model both include two hidden layers, each containing 128 neurons. Moreover, the activation function for the critic network is Mish, and the activation function for the final layer of the denoising network is Tanh. The size of experience replay buffer and mini-batch are set to $10^{6}$ and 128, respectively. Moreover, the learning rate is configured to 0.0003.
\end{itemize}

\subsection{Performance Analysis}
\par
\begin{figure}
    \centering
    \includegraphics[width=\linewidth]{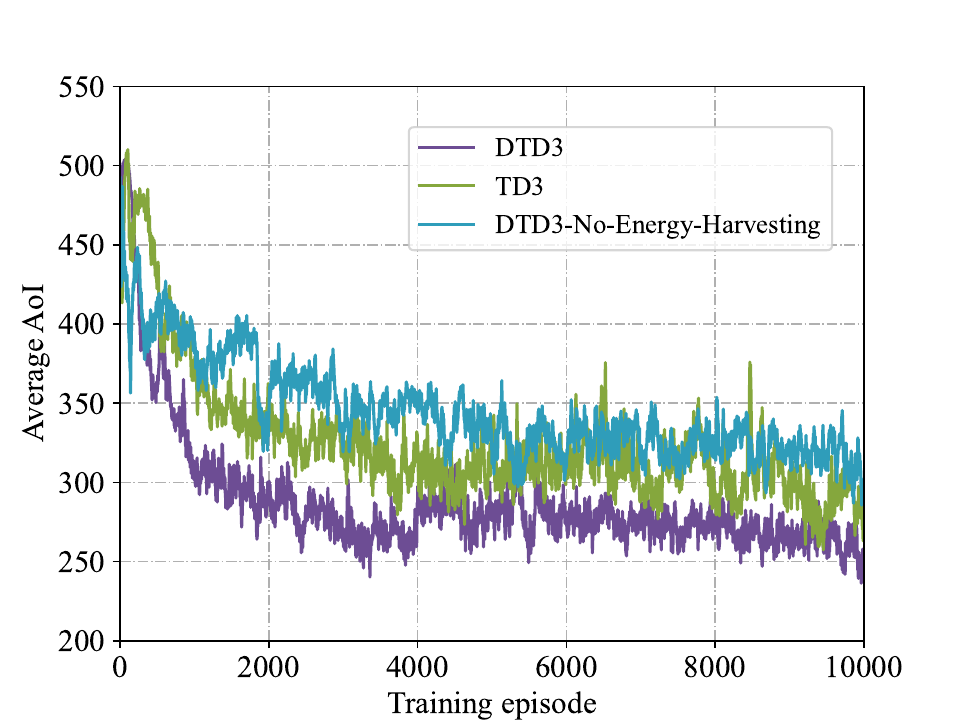}
    \caption{The AoI convergence of different algorithms.}
    \label{fig: AoI}
\end{figure}

 The AoI convergence curves for various algorithms are illustrated in Fig.~\ref{fig: AoI}. We can find that the AoI of IoT devices becomes higher when the UAV do not provide energy transfer services. This is due to the fact that uploading data consumes energy, and IoT devices have limited energy storage. Therefore, when the energy buffer becomes empty, they are unable to continue uploading data, leading to an increased AoI. Moreover, we can observe that the proposed DTD3 algorithm achieves better AoI performance and converges more rapidly and stably than the traditional TD3 algorithm. This performance improvement can be attributed to the diffusion model. First, the analytical ability of diffusion model allows for a better exploration of the latent relationship between states and actions. Second, by leveraging the generative and data augmentation abilities of the diffusion model, DTD3 can produce diverse high-quality actions, thus improving the exploration of the action space compared to the traditional TD3.



\section{Future Directions}
\par In this section, we present future directions about GenAI-powered energy harvesting wireless network.

\subsection{GenAI-based Distributed Energy Beamforming}
\par Distributed energy beamforming allows multiple dispersed energy sources to work together like an antenna array by simultaneously directing RF energy towards a specific energy harvester. In this case, GenAI can perform coordination of the distributed carriers in phase and frequency so that the RF signals can be constructively combined at the receiver.

\subsection{GenAI-based Inference Management}
\par Interference is a common factor affecting wireless networks. Currently, most research focuses on how to avoid and mitigate interference. In energy-harvesting wireless networks, leveraging GenAI to devise intelligent scheduling strategies that transform harmful interference into useful energy represents a promising and intriguing direction for future exploration.

\subsection{GenAI-based Trade-off Strategy Design}
\par For RF-powered devices, where the transmission power is generally low, employing multiple antennas can enhance transmission efficiency. However, increasing the number of antennas also leads to higher power consumption. GenAI can be utilized to develop suitable optimization strategies that balance transmission efficiency and power consumption, making it well-suited for dynamic environments.

\section{Conclusion}

\par In this article, we have introduced how GenAI optimizes the energy harvesting wireless network. Specifically, we have first presented the fundamentals of GenAI, different types of energy source, and the architecture of energy harvesting network. Then, we have discussed several GenAI-based methods to address various issues of energy harvesting wireless network. Subsequently, we have conducted a case study on UAV-enabled energy transfer and data collection to demonstrate the importance of energy harvesting technology in the IoT network and validate the effectiveness of GenAI. Finally, three key future directions have been shown that can further improve GenAI on energy harvesting wireless network. We hope that this article can inspire researchers to propose more GenAI-based methods in IoT network.





\bibliography{main}

\begin{thebibliography}{10}
\providecommand{\url}[1]{#1}
\csname url@samestyle\endcsname
\providecommand{\newblock}{\relax}
\providecommand{\bibinfo}[2]{#2}
\providecommand{\BIBentrySTDinterwordspacing}{\spaceskip=0pt\relax}
\providecommand{\BIBentryALTinterwordstretchfactor}{4}
\providecommand{\BIBentryALTinterwordspacing}{\spaceskip=\fontdimen2\font plus
\BIBentryALTinterwordstretchfactor\fontdimen3\font minus \fontdimen4\font\relax}
\providecommand{\BIBforeignlanguage}[2]{{%
\expandafter\ifx\csname l@#1\endcsname\relax
\typeout{** WARNING: IEEEtran.bst: No hyphenation pattern has been}%
\typeout{** loaded for the language `#1'. Using the pattern for}%
\typeout{** the default language instead.}%
\else
\language=\csname l@#1\endcsname
\fi
#2}}
\providecommand{\BIBdecl}{\relax}
\BIBdecl

\bibitem{Khanna2020}
A.~Khanna and S.~Kaur, ``Internet of things (iot), applications and challenges: {A} comprehensive review,'' \emph{Wirel. Pers. Commun.}, vol. 114, no.~2, pp. 1687--1762, 2020.

\bibitem{Lu2015}
X.~Lu, P.~Wang, D.~Niyato, D.~I. Kim, and Z.~Han, ``Wireless networks with {RF} energy harvesting: {A} contemporary survey,'' \emph{{IEEE} Commun. Surv. Tutorials}, vol.~17, no.~2, pp. 757--789, 2015.

\bibitem{Liu2022}
L.~Liu, K.~Xiong, J.~Cao, Y.~Lu, P.~Fan, and K.~B. Letaief, ``Average {A}o{I} minimization in {UAV}-assisted data collection with {RF} wireless power transfer: {A} deep reinforcement learning scheme,'' \emph{{IEEE} Internet Things J.}, vol.~9, no.~7, pp. 5216--5228, 2022.

\bibitem{nando2024enhancing}
Y.~A. Nando and W.-Y. Chung, ``Enhancing rf energy harvesting and wireless power transfer with gan-optimized 3d quasi-yagi antenna,'' in \emph{{WPTCE}}, 2024, pp. 454--458.

\bibitem{zargari2024deep}
S.~Zargari, C.~Tellambura, A.~Maaref, and G.~Y. Li, ``Deep conditional generative adversarial networks for efficient channel estimation in ambc systems,'' \emph{IEEE Transactions on Machine Learning in Communications and Networking}, 2024.

\bibitem{ahmadi2019renewable}
M.~H. Ahmadi, M.~Ghazvini, M.~Alhuyi~Nazari, M.~A. Ahmadi, F.~Pourfayaz, G.~Lorenzini, and T.~Ming, ``Renewable energy harvesting with the application of nanotechnology: A review,'' \emph{International Journal of Energy Research}, vol.~43, no.~4, pp. 1387--1410, 2019.

\bibitem{Sah2020}
D.~K. Sah and T.~Amgoth, ``Renewable energy harvesting schemes in wireless sensor networks: {A} survey,'' \emph{Inf. Fusion}, vol.~63, pp. 223--247, 2020.

\bibitem{Guo2018}
S.~Guo, Y.~Shi, Y.~Yang, and B.~Xiao, ``Energy efficiency maximization in mobile wireless energy harvesting sensor networks,'' \emph{{IEEE} Trans. Mob. Comput.}, vol.~17, no.~7, pp. 1524--1537, 2018.

\bibitem{8322138}
L.~Yang, Y.~J. Zhou, C.~Zhang, X.~M. Yang, X.-X. Yang, and C.~Tan, ``Compact multiband wireless energy harvesting based battery-free body area networks sensor for mobile healthcare,'' \emph{IEEE Journal of Electromagnetics, RF and Microwaves in Medicine and Biology}, vol.~2, no.~2, pp. 109--115, 2018.

\bibitem{6206429}
A.~Georgiadis and A.~Collado, ``Improving range of passive {RFID} tags utilizing energy harvesting and high efficiency class-e oscillators,'' in \emph{{EUCAP}}, 2012, pp. 3455--3458.

\bibitem{Chung2023}
K.~Chung and J.~Lim, ``Machine learning for relaying topology: Optimization of {I}o{T} networks with energy harvesting,'' \emph{{IEEE} Access}, vol.~11, pp. 41\,827--41\,839, 2023.

\bibitem{Lin2024}
R.~Lin, H.~Qiu, J.~Wang, Z.~Zhang, L.~Wu, and F.~Shu, ``Physical-layer security enhancement in energy-harvesting-based cognitive internet of things: {A} {GAN}-powered deep reinforcement learning approach,'' \emph{{IEEE} Internet Things J.}, vol.~11, no.~3, pp. 4899--4913, 2024.

\bibitem{kaur2023vae}
D.~Kaur, S.~N. Islam, M.~A. Mahmud, M.~E. Haque, and A.~Anwar, ``A {VAE}-bayesian deep learning scheme for solar power generation forecasting based on dimensionality reduction,'' \emph{Energy and AI}, vol.~14, p. 100279, 2023.

\bibitem{Hu2021}
H.~Hu, K.~Xiong, G.~Qu, Q.~Ni, P.~Fan, and K.~B. Letaief, ``{A}o{I}-minimal trajectory planning and data collection in {UAV}-assisted wireless powered {I}o{T} networks,'' \emph{{IEEE} Internet Things J.}, vol.~8, no.~2, pp. 1211--1223, 2021.

\bibitem{Fujimoto2018}
S.~Fujimoto, H.~van Hoof, and D.~Meger, ``Addressing function approximation error in actor-critic methods,'' in \emph{Proceedings of the 35th International Conference on Machine Learning, {ICML}}, vol.~80, 2018, pp. 1582--1591.

\end{thebibliography}

\end{document}

